\documentclass[12pt,titlepage]{article}
\usepackage[dvips]{graphicx}

\begin{document}

\title{Exact closed-form solutions of the Dirac equation with a scalar exponential
potential}
\author{Antonio S. de Castro\thanks{%
castro@feg.unesp.br} and Marcelo Hott\thanks{%
hott@feg.unesp.br} \\
\\
UNESP - Campus de Guaratinguet\'{a}\\
Departamento de F\'{\i}sica e Qu\'{\i}mica\\
Caixa Postal 205\\
12516-410 Guaratinguet\'{a} SP - Brasil}
\date{}
\maketitle

\begin{abstract}
The problem of a fermion subject to a general scalar potential in a
two-dimensional world for nonzero eigenenergies is mapped into a
Sturm-Liouville problem for the upper component of the Dirac spinor. In the
specific circumstance of an exponential potential, we have an effective
Morse potential which reveals itself as an essentially relativistic problem.
Exact bound solutions are found in closed form for this problem. The
behaviour of the upper and lower components of the Dirac spinor is discussed
in detail, particularly the existence of zero modes.
\end{abstract}

The Coulomb potential of a point electric charge in a 1+1 dimension,
considered as the time component of a Lorentz vector, is linear ($\sim x$)
and so it provides a constant electric field always pointing to, or from,
the point charge. This problem is related to the confinement of fermions in
the Schwinger and in the massive Schwinger models \cite{col1}-\cite{col2}
and in the Thirring-Schwinger model \cite{fro}. It is frustrating that, due
to the tunneling effect (Klein\'{}s paradox), there are no bound states for
this kind of potential regardless of the strength of the potential \cite{cap}%
-\cite{gal}. The linear potential, considered as a Lorentz scalar, is also
related to the quarkonium model in one-plus-one dimensions \cite{hoo}-\cite
{kog}. Recently it was incorrectly concluded that even in this case there is
solely one bound state \cite{bha}. Later, the proper solutions for this last
problem were found \cite{cas1}-\cite{hil}. The mixed vector-scalar potential
has also been analyzed for a linear potential \cite{cas2} as well as for a
general potential which goes to infinity as $x\rightarrow \infty $ \cite{ntd}%
. In both of those last references it has been concluded that there is
confinement if the scalar coupling is of sufficient intensity compared to
the vector coupling.

The problem of a particle subject to an inversely linear potential in one
spatial dimension ($\sim |x|^{-1}$), known as the one-dimensional hydrogen
atom, has also received considerable attention in the literature. This
problem presents some conundrums and the most perplexing one is regarding to
the ground state. The nonrelativistic Schr\"{o}dinger equation provides a
ground-state solution with infinite eigenenergy and a related eigenfunction
given by a delta function centered about the origin. This problem was also
analyzed with the Klein-Gordon equation and there it was revealed a finite
eigenenergy and an exponentially decreasing eigenfunction \cite{spe}. The
problem was also approached with the Schr\"{o}dinger, Klein-Gordon and Dirac
equations and it was concluded that the Klein-Gordon equation provides
unacceptable solutions while the Dirac equation, with the interacting
potential considered as a time component of a vector, has no bounded
solutions at all \cite{mos}. This problem was also sketched for a Lorentz
scalar interacting potential in the Dirac equation \cite{ho}, but the
analysis is incomplete. In a recent paper \cite{pl}, it was shown that the
problem of a fermion under the influence of a general scalar potential for
nonzero eigenenergies can be mapped into a Sturm-Liouville problem. Next,
the key conditions for the existence of bound-state solutions were settled
for power-law potentials. In addition, the solution for an inversely linear
potential, including the zero-eigenmode, was obtained in closed form.

The problem of a fermion subject to a well potential given either by an
exponential function ($V=V_{0}\exp (-ar)$) is exactly solved for S-wave
bound-states in the nonrelativistic quantum mechanics \cite{flu}. On the
other side, the problem for S-states for the Morse potential ($V=V_{0}\left[
1-\exp (-ar)\right] ^{2}$) is transformed into the difficult problem of
solving a transcendent equation which is only approximately solvable \cite
{flu}-\cite{haa1}. The first problem presents some interest to the
nucleon-nucleon system due to the short-range character of the interaction
\cite{br}-\cite{bm}, whereas the second one has been used to describe the
vibrations of nuclei in homonuclear diatomic molecules \cite{mor1}-\cite
{mor2}. The one-dimensional asymmetric Morse potential ($-\infty <r<\infty $%
) is also an exactly solvable problem in the nonrelativistic quantum
mechanics \cite{haa}-\cite{nie}, even if its parameters are complex numbers
\cite{ahmed}. In the present paper we approach the time-independent Dirac
equation in 1+1 dimensions with a scalar potential given by an exponential
potential. For nonzero eigenenergies this sort of potential gives rise to an
effective Morse potential in a Sturm-Liouville problem for the upper
component of the Dirac spinor. We find the bound states and discuss the
existence of zero modes, which are related to the ultrarelativistic limit of
the Dirac equation and play an important role in the induction of a
fractional fermion number on the vacuum in the second quantization setting.

The two-dimensional Dirac equation can be obtained from the
four-dimen\-sional one with the mixture of spherically symmetric scalar,
vector and anomalous magnetic-like (tensor) interactions. If we limit the
fermion to move in the $x$-direction ($p_{y}=p_{z}=0$) the four-dimensional
Dirac equation decomposes into two equivalent two-dimensional equations with
2-component spinors and 2$\times $2 matrices \cite{str}. Then, there results
that the scalar and vector interactions preserve their Lorentz structures
whereas the anomalous magnetic interaction turns out to be a pseudoscalar
interaction. Furthermore, in the 1+1 world there is no angular momentum so
that the spin is absent. Therefore, the 1+1 dimensional Dirac equation allow
us to explore the physical consequences of the negative-energy states in a
mathematically simpler and more physically transparent way.

In the presence of a time-independent scalar potential the 1+1 dimensional
time-independent Dirac equation for a fermion of rest mass $m$ reads

\begin{equation}
\left[ c\alpha p+\beta \left( mc^{2}+V\right) \right] \psi =E\psi ,
\label{1}
\end{equation}

\noindent where $E$ is the energy of the fermion, $c$ is the velocity of
light and $p$ is the momentum operator. $\alpha $ and $\beta $ are Hermitian
square matrices satisfying the relations $\alpha ^{2}=\beta ^{2}=1$, $%
\left\{ \alpha ,\beta \right\} =0$. One can choose the 2$\times $2 Pauli
matrices satisfying the same algebra as $\alpha $ and $\beta $, resulting in
a 2-component spinor $\psi $. The positive definite function $|\psi
|^{2}=\psi ^{\dagger }\psi $, satisfying a continuity equation, is
interpreted as a position probability density and its norm is a constant of
motion. This interpretation is completely satisfactory for single-particle
states \cite{tha}. Using $\alpha =\sigma _{2}$, $\beta =\sigma _{1}$ and
provided that the spinor is written in terms of the upper and the lower
components
\begin{equation}
\psi =\left(
\begin{array}{c}
\phi \\
\chi
\end{array}
\right) ,  \label{2}
\end{equation}

\noindent the Dirac equation decomposes into :

\begin{eqnarray}
E\phi &=&\left( mc^{2}+V\right) \chi -\hbar c\chi ^{\prime },  \label{3a} \\
&&  \nonumber \\
E\chi &=&\left( mc^{2}+V\right) \phi +\hbar c\phi ^{\prime },  \label{3b}
\end{eqnarray}

\noindent where the prime denotes differentiation with respect to $x$. In
terms of $\phi $ and $\chi $ the spinor is normalized as $\int_{-\infty
}^{+\infty }dx\left( |\phi |^{2}+|\chi |^{2}\right) =1$, so that $\phi $ and
$\chi $ are square integrable functions. It is remarkable that the Dirac
equation with a scalar potential is not invariant under $V\rightarrow
V+const.$ Therefore, the absolute values of the energy will have physical
significance and the freedom to choose a zero-energy will be lost.

Furthermore, using the expression for $\chi $ obtained from (\ref{3b}), viz.
\begin{equation}
\chi =\frac{\left( mc^{2}+V\right) \phi +\hbar c\,\phi ^{\prime }}{E},\ \
E\neq 0,  \label{8a}
\end{equation}

\noindent and inserting it in (\ref{3a}) one arrives at the following
second-order differential equation for $\phi $:

\begin{equation}
-\frac{\hbar ^{2}}{2}\;\phi ^{\prime \prime }+V_{eff}\;\phi =E_{eff}\;\phi ,
\label{8}
\end{equation}

\noindent where
\begin{eqnarray}
E_{eff} &=&\frac{E^{2}-m^{2}c^{4}}{2c^{2}}  \label{9} \\
&&  \nonumber \\
V_{eff} &=&\frac{V^{2}}{2c^{2}}+mV-\frac{\hbar }{2c}V^{\prime }.  \label{10}
\end{eqnarray}

\noindent \noindent \noindent \noindent \noindent The energy levels are
symmetrical about $E=0$ (see, \textit{e.g.}, Refs. \cite{cn} and \cite{cnt}%
). This conclusion can be obtained directly from (\ref{9}). Besides, if $%
\psi $ is a solution with energy $E$ then $\sigma _{3}\psi ^{*}$ is also a
solution with energy $-E$ for the very same potential. It means that the
potential couples to the positive-energy component of the spinor in the same
way it couples to the negative-energy component. In other words, this sort
of potential couples to the mass of the fermion instead of its charge so
that there is no atmosphere for the production of particle-antiparticle
pairs. No matter the intensity and sign of the coupling parameter, the
positive- and the negative-energy solutions never meet. Thus there is no
room for transitions from positive- to negative-energy solutions. This all
means that Klein\'{}s paradox never comes to the scenario.

Up to this point we have considered solutions for $E\neq 0$. Nevertheless,
one could also ask for possible zero-energy solutions. These zero-mode
solutions can be obtained directly from the Dirac equation (\ref{3a})-(\ref
{3b}). One can observe that the Dirac Hamiltonian with a general scalar
potential always supports a zero-mode solution with upper and lower
components given by
\begin{eqnarray}
\phi &=&N_{\phi }\exp \left[ -\left( \frac{mc^{2}x+v(x)}{\hbar c}\right)
\right]  \label{29a} \\
&&  \nonumber \\
\chi &=&N_{\chi }\exp \left[ +\left( \frac{mc^{2}x+v(x)}{\hbar c}\right)
\right] ,  \label{29b}
\end{eqnarray}

\noindent where $N_{\phi }$ and $N_{\chi }$ are constants and

\begin{equation}
v(x)=\int^{x}V(y)\,dy.  \label{29c}
\end{equation}

\noindent One can check that it is impossible to have both components
different from zero simultaneously as physically acceptable solutions. A
normalizable zero-mode eigenstate with a nonzero upper component for a
massive fermion is possible only if $v(x)$ has a leading asymptotic
behaviour given by

\begin{equation}
\lim_{x\rightarrow +\infty }v(x)\sim \left\{
\begin{array}{c}
{\textrm{constant\thinspace }} \\
\mathrm{or} \\
+\infty \\
\mathrm{or} \\
\pm |x|^{1-\delta }\,,\,\,\,\,0<\delta <1 \\
\mathrm{or} \\
-\Delta |x|,\,-\infty \,<\Delta <mc^{2} \\
\mathrm{or} \\
-k\ln |x|,\,\,(k>0)
\end{array}
\,\right.  \label{29d}
\end{equation}

\noindent and

\begin{equation}
\lim_{x\rightarrow -\infty }v(x)\sim \left\{
\begin{array}{c}
|x|^{1+\varepsilon },\,\mathrm{\,}\varepsilon >0 \\
\mathrm{or} \\
+\Delta |x|,\,\,\Delta >mc^{2}.
\end{array}
\right.  \label{29e}
\end{equation}

\noindent On the other hand, a normalizable zero-mode eigenstate with a
nonzero lower component for a massive fermion is possible only if one of the
following restrictions below is satisfied

\begin{equation}
\lim_{x\rightarrow +\infty }v(x)\sim \left\{
\begin{array}{c}
-|x|^{1+\varepsilon },\,\mathrm{\,}\varepsilon >0 \\
\mathrm{or} \\
-\Delta |x|,\,\,\,\mathrm{\,}\Delta >mc^{2},
\end{array}
\right.  \label{29f}
\end{equation}

\noindent and

\begin{equation}
\lim_{x\rightarrow -\infty }v(x)\sim \left\{
\begin{array}{c}
{\textrm{constant\thinspace }} \\
\mathrm{or} \\
-\infty \\
\mathrm{or} \\
\pm |x|^{1-\delta }\,,\,\,\,\,0<\delta <1 \\
\mathrm{or} \\
+\Delta |x|,\,-\infty \,<\Delta <mc^{2} \\
\mathrm{or} \\
+k\ln |x|,\,\,(k>0).
\end{array}
\right.  \label{29g}
\end{equation}

\noindent Notice that for a massless fermion it is enough that $v(x)$ grows
faster than $\hbar c|x|$ $(-\hbar c|x|)$ as $|x|\rightarrow \infty $ to have
a nonzero upper (lower) component.

Incidentally, the existence of Dirac eigenspinors with a vanishing lower
component, or even with a vanishing upper component, is due to the
particular representations of the matrices $\alpha $ and $\beta $ adopted in
this paper. It is instructive at this point to consider for a moment a
representation where the eigenspinor presents a more familiar behaviour. Let
us write the Dirac equation (\ref{1}) as
\begin{equation}
\left[ c\sigma _{1}p+\sigma _{3}\left( mc^{2}+V\right) \right] \tilde{\psi}=E%
\tilde{\psi}.  \label{29}
\end{equation}

\noindent The original spinor is related to $\tilde{\psi}$ by the unitary
transformation $\psi =U\tilde{\psi}$, where
\begin{equation}
U=\frac{1}{\sqrt{2}}\left(
\begin{array}{ll}
1 & -i \\
1 & +i
\end{array}
\right) ,  \label{30}
\end{equation}

\noindent so that $\phi =\left( \tilde{\phi}-i\tilde{\chi}\right) /\sqrt{2}$
and $\chi =\left( \tilde{\phi}+i\tilde{\chi}\right) /\sqrt{2}$. In the
nonrelativistic approximation (\ref{29}) becomes

\begin{equation}
\tilde{\chi}=\frac{p}{2mc}\;\tilde{\phi}  \label{eq8c}
\end{equation}

\begin{equation}
\left( -\frac{\hbar ^{2}}{2m}\frac{d^{2}}{dx^{2}}+V\right) \tilde{\phi}%
=\left( E-mc^{2}\right) \tilde{\phi}.  \label{eq8d}
\end{equation}

\noindent Eq. (\ref{eq8c}) shows that $\tilde{\chi}$ is of order $v/c<<1$
relative to $\tilde{\phi}$ and Eq. (\ref{eq8d}) shows that $\tilde{\phi}$
obeys the Schr\"{o}dinger equation with the potential $V$. Now one can see
that when one uses the representation where $\alpha =\sigma _{2}$, $\beta
=\sigma _{1}$ (that one used in this paper) one obtains upper and lower
components approximately equal to each other in the nonrelativistic limit.
On the other side, in the ultrarelativistic limit one expects that $\tilde{%
\chi}$ presents a contribution comparable to $\tilde{\phi}$, thus the
possibilities $\tilde{\phi}\approx -i\tilde{\chi}$ and $\tilde{\phi}\approx
+i\tilde{\chi}$ imply into $\chi \approx 0$ and $\phi \approx 0$,
respectively. Therefore, one can conclude that the zero-mode solutions
correspond to the ultrarelativistic limit of the theory.

Now let us focus our attention on a nonconserving-parity scalar potential in
the form
\begin{equation}
V(x)=A\exp \left( -\frac{\lambda }{\hbar c}x\right) ,  \label{11}
\end{equation}
\noindent where $\lambda $ is a positive parameter related to the range of
the interaction.

It follows from Eqs. (\ref{29d})-(\ref{29g}) that only when $A<0$, this
potential meets the requirement to hold a zero-eigenmode. In passing, the
lower component of the Dirac spinor vanishes. In terms of $\xi $ and $\mu $
defined by

\negthinspace
\begin{eqnarray}
\xi &=&2\frac{|A|}{\lambda }\exp \left( -\frac{\lambda }{\hbar c}x\right)
\qquad  \nonumber \\
&&  \label{11a} \\
\mu &=&\frac{\sqrt{m^{2}c^{4}-E^{2}}}{\lambda },  \nonumber
\end{eqnarray}

\noindent the normalizable Dirac spinor corresponding to $E=0$ is, then

\begin{equation}
\psi =N\,\xi ^{\mu }\,e^{-\xi /2}\left(
\begin{array}{l}
1 \\
0
\end{array}
\right) ,  \label{28}
\end{equation}

\noindent where $N$ is a normalization constant. Note that the normalization
of this spinor is possible only if $\mu >0$, then the fermions must be
massive. Once this criterion is verified the zero-eigenmode solution always
exists for any negative value of $A$, regardless the mass of the fermion.

For $E\neq 0$, the potential (\ref{11}) gives rise to the effective
potential
\begin{equation}
V_{eff}(x)=V_{1}\exp \left( -2\frac{\lambda }{\hbar c}x\right) +V_{2}\exp
\left( -\frac{\lambda }{\hbar c}x\right) ,  \label{12}
\end{equation}
\noindent where
\begin{eqnarray}
V_{1} &=&\frac{A^{2}}{2c^{2}}>0  \nonumber \\
&&  \label{13} \\
V_{2} &=&\frac{\lambda A}{c^{2}}\left( \frac{mc^{2}}{\lambda }+\frac{1}{2}%
\right) .  \nonumber
\end{eqnarray}

\noindent The effective potential tends to $+\infty $ as $x\rightarrow
-\infty $ and vanishes as $x\rightarrow +\infty $. Furthermore there is a
local minimum if $V_{2}<0$. This condition for the existence of a local
minimum requires that $A<0$. The effective Morse potential with a well
structure might hold bound states in spite of the fact that the original
exponential potential given by (\ref{11}) is everywhere repulsive, so that
one can not expect bound solutions in the nonrelativistic limit. This
essentially relativistic solution is due to the peculiar coupling in the
Dirac equation: the scalar potential behaves like an $x$-dependent rest mass
\cite{tha}. Using (\ref{8})-(\ref{10}) one obtains the equation

\begin{equation}
\xi \phi ^{\prime \prime }+\phi ^{\prime }+\left( -\frac{\xi }{4}-\frac{\mu
^{2}}{\xi }+\rho \right) \phi =0,  \label{16}
\end{equation}

\noindent where $\rho $ is defined by
\[
\rho =\frac{mc^{2}}{\lambda }+\frac{1}{2},
\]

\noindent and the prime denotes differentiation with respect to $\xi $. The
normalizable asymptotic form of the solution as $\xi \rightarrow \infty $ is
$e^{-\xi /2}$. As $\xi \rightarrow 0$, when the term $1/\xi $ dominates, the
regular solution behaves as $\xi ^{\mu }$ and the solution for all $\xi $
can be expressed as $\phi (\xi )=\xi ^{\mu }e^{-\xi /2}w(\xi )$, where $w$
is solution of the confluent hypergeometric equation \cite{abr}

\begin{equation}
\xi w^{\prime \prime }+(b-\xi )w^{\prime }-aw=0,  \label{19}
\end{equation}

\noindent with
\begin{eqnarray}
a &=&\frac{b}{2}-\rho  \nonumber \\
&&  \label{20} \\
b &=&2\mu +1.  \nonumber
\end{eqnarray}

\noindent Then $w$ is expressed as $_{1\!\;}\!F_{1}(a,b,\xi )$ and in order
to furnish normalizable $\phi $, the confluent hypergeometric function must
be a polynomial. This demands that $a=-n$, where $n$ is a nonnegative
integer in such a way that
\begin{equation}
\mu =\frac{mc^{2}}{\lambda }-n,  \label{20a}
\end{equation}
\noindent and $_{1\!\;}\!F_{1}(a,b,\xi )$ is proportional to the associated
Laguerre polynomial $L_{n}^{2\mu }(\xi )$, a polynomial of degree $n$. This
requirement, combined with (\ref{11a}) and (\ref{20}), implies into
quantized energy eigenvalues:
\begin{equation}
E_{n}=\pm \lambda \sqrt{n\left( 2\mu +n\right) \,}  \label{20b}
\end{equation}
Finally, we are ready to present the solution for the upper component of the
Dirac spinor:

\begin{equation}
\phi _{n}=N\xi ^{\mu }e_{\;}^{-\xi /2}\;L_{n}^{2\mu }\left( \xi \right) ,
\label{23}
\end{equation}
The solution for the lower component, obtained by inserting (\ref{23}) into (%
\ref{8a}) and using recurrence relations among the associated Laguerre
polynomials \cite{abr}, is given by

\begin{equation}
\chi _{n}=\pm \sqrt{\frac{2\mu +n}{n}}\,N\xi ^{\mu }e_{\;}^{-\xi
/2}L_{n-1}^{2\mu }\left( \xi \right) .  \label{23b}
\end{equation}

\noindent Since $\mu >0$, one can conclude from (\ref{20a}) that there is a
finite number of bound-state solutions. Further, the components of the Dirac
spinor has a nodal structure in such a way that the number of nodes of the
lower component is given by $n-1$. These facts imply that the quantum number
$n$, for $E\neq 0$, is given by $n=1,2,3,\ldots <mc^{2}/\lambda $. \noindent
Note that $\lambda <mc^{2}$ in order that the potential holds at least one
excited-state solution.

Therefore, the solution of the problem, including the zero-eigenmode
corresponding to $n=0$, can be written more succinctly as
\begin{eqnarray}
E_{n} &=&\pm \lambda \sqrt{n\left( 2\frac{mc^{2}}{\lambda }-n\right) \,},\ \
n=0,1,2,\ldots <mc^{2}/\lambda  \nonumber \\
&&  \label{27} \\
\psi _{n} &=&N_{n}\xi ^{\mu }e_{\;}^{-\xi /2}\left(
\begin{array}{l}
L_{n}^{2\mu }\left( \xi \right) \\
\\
R_{n}\,L_{n-1}^{2\mu }\left( \xi \right)
\end{array}
\right) ,  \nonumber
\end{eqnarray}

\noindent where
\begin{equation}
R_{n}=\left\{
\begin{array}{l}
0 \\
\\
\pm \sqrt{\frac{2\mu +n}{n}}
\end{array}
\begin{array}{l}
,\hspace{0.25cm}\mathrm{for}\hspace{0.25cm}n=0 \\
\\
,\hspace{0.25cm}\mathrm{for}\hspace{0.25cm}n\neq 0.
\end{array}
\right.
\end{equation}

\noindent The top line of Eq. (\ref{27}), implies that the Dirac energy
eigenvalues are restricted to the range $|E|<mc^{2}$. The energies belonging
to $|E|>mc^{2}$ correspond to the continuum. This conclusion could also be
obtained from the definition of $\mu $ in (\ref{11a}). In order to evaluate
the normalization constant $N_{n}$ we are confronted by the integral
\begin{equation}
I_{n}=\int_{0}^{\infty }d\xi \,\xi ^{2\mu -1}e^{-\xi }\left( L_{n}^{2\mu
}\right) ^{2}.  \label{27a}
\end{equation}

\noindent One might think that the singularity of the integrand in (\ref{27a}%
) could be a source of embarrassment. However, the integral is convergent
for $\mu >0$, as it is the case in hands, although it is not so easy
calculate it. Fortunately, it has been already done in Refs. \cite{nie}-\cite
{ahmed}. There results that

\begin{equation}
I_{n}=\frac{\Gamma \left( 2\mu +n+1\right) }{2\mu \,\Gamma \left( n+1\right)
}.  \label{27b}
\end{equation}

\noindent It takes a little algebra to conclude that the normalization
constant can be compactly written as
\begin{equation}
N_{n}=\sqrt{\frac{\lambda /\left( \hbar c\right) }{I_{n}+R_{n}^{2}I_{n-1}}}.
\label{27c}
\end{equation}

\noindent

\begin{figure}[!ht]
\begin{center}
\includegraphics[width=8cm, angle=270]{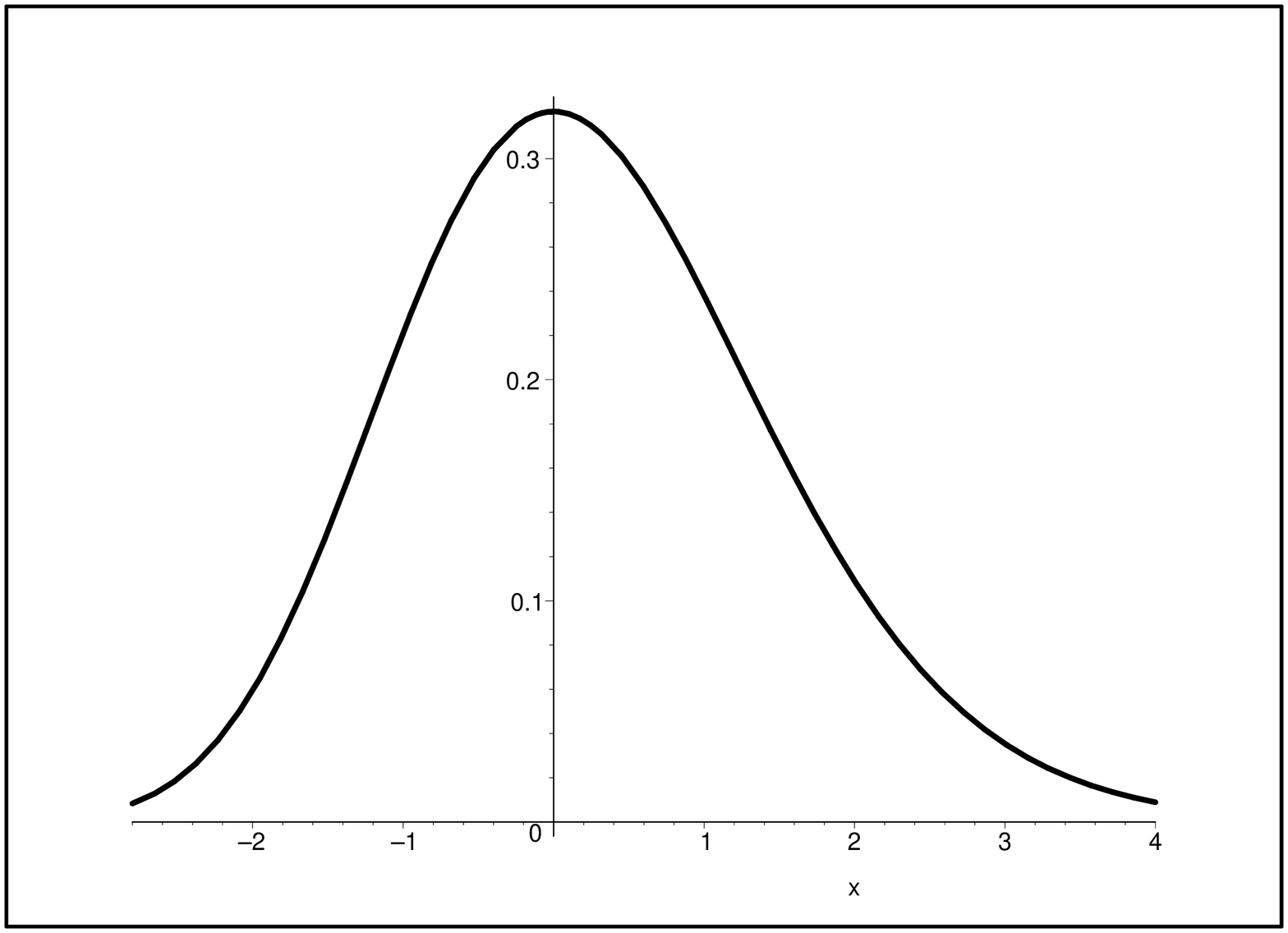}
\end{center}
\par
\vspace*{-0.1cm}
\caption{$|\psi |^{2}=|\phi |^{2}$ as a function of $x$ corresponding to the
ground state ($E=0$) of the potential $V(x)=A\exp \left( -\frac{\lambda }{%
\hbar c}x\right)$ ($m=\hbar =c=1,\,A=-1$ and $\lambda= 1/3$).}
\label{Fig1}
\end{figure}

\begin{figure}[!ht]
\begin{center}
\includegraphics[width=8cm, angle=270]{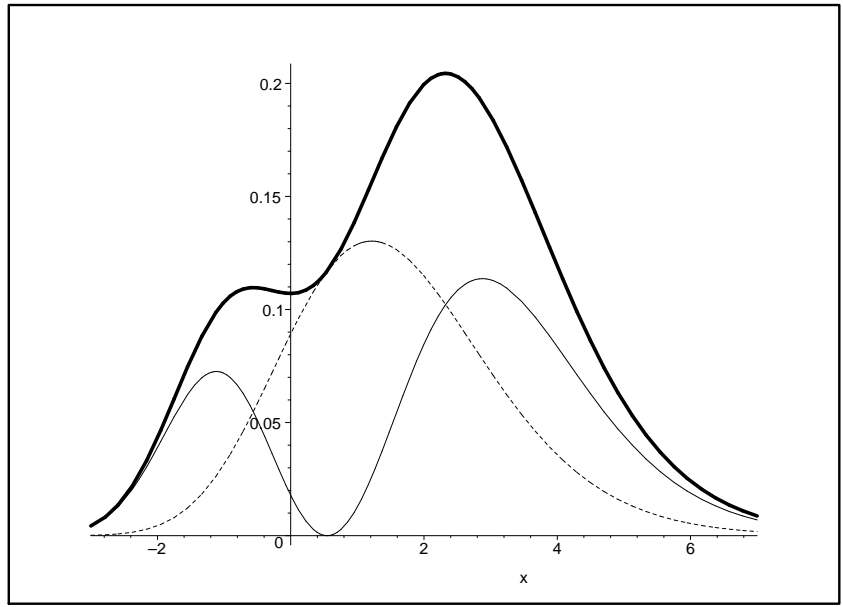}
\end{center}
\par
\vspace*{-0.1cm}
\caption{$|\phi |^{2}$ (full thin line), $|\chi |^{2}$ (dashed line) and $%
|\psi |^{2}=|\phi |^{2}+|\chi |^{2}$ (full thick line) as a function of $x$
for the first-excited state of the potential $V(x)=A\exp \left( -\frac{%
\lambda }{\hbar c}x\right)$($m=\hbar =c=1,\,A=-1$ and $\lambda= 1/3$).}
\label{Fig2}
\end{figure}

\begin{figure}[!ht]
\begin{center}
\includegraphics[width=8cm, angle=270]{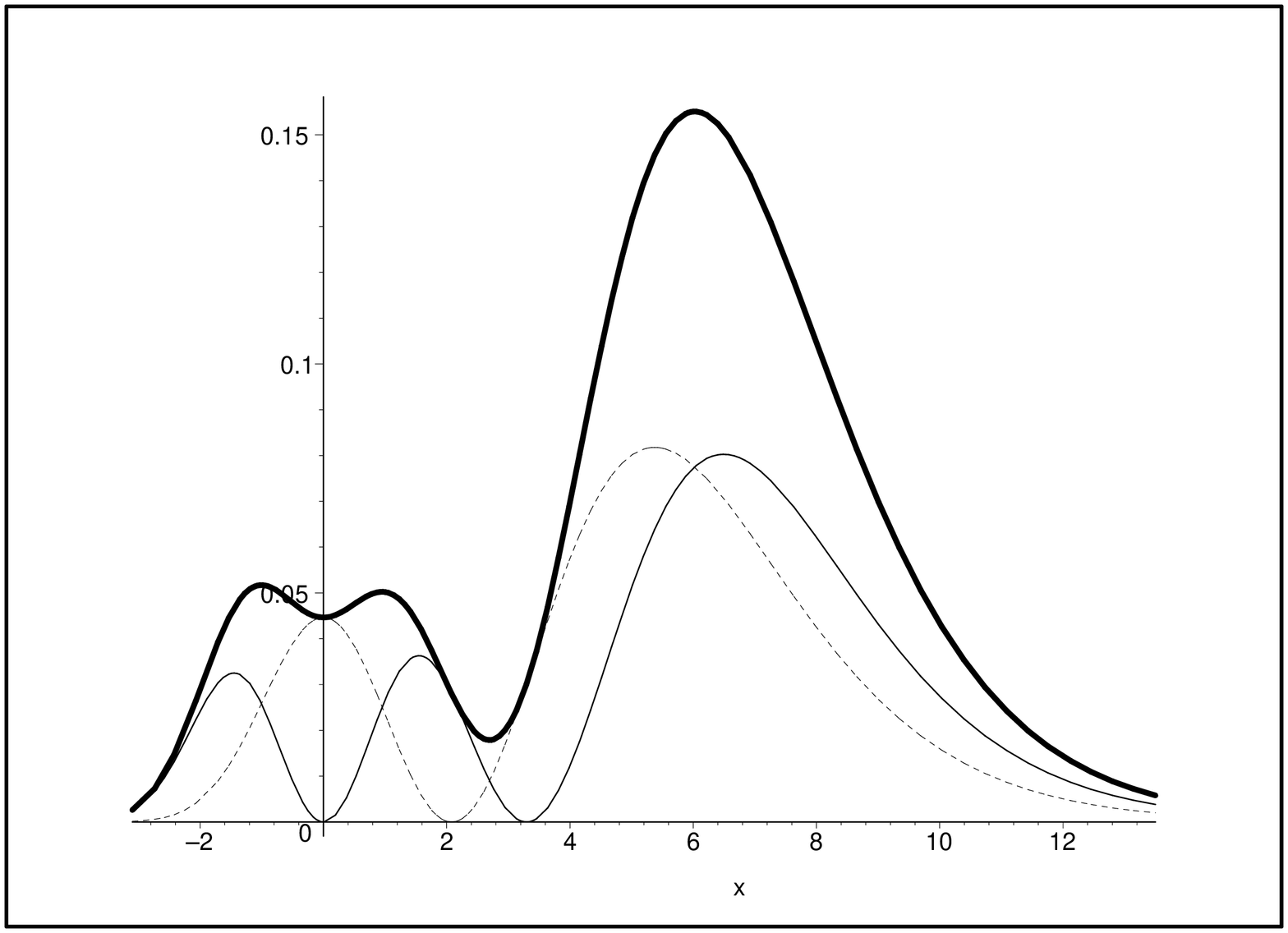}
\end{center}
\par
\vspace*{-0.1cm}
\caption{The same as in Fig. 2 for the second-excited state.}
\label{Fig3}
\end{figure}

In order to get further confidence on the normalizability of the Dirac
eigenspinor as well as to get a better understanding of its behaviour , a
few plots for $|\phi |^{2}$ and $|\chi |^{2}$ are presented. Fig. \ref{Fig1}
illustrates the behaviour of the position probability density, $|\psi
|^{2}=|\phi |^{2}$, for the zero-mode solution. Figs. \ref{Fig2} and \ref
{Fig3} illustrate the behaviour of the upper and lower components of the
Dirac spinor, $|\phi |^{2}$ and $|\chi |^{2}$, and the position probability
density, $|\psi |^{2}=|\phi |^{2}+|\chi |^{2}$, for the positive-energy
solutions of the first- and second-excited states, respectively. The results
for negative energies are the same as far as the charge conjugation does $%
\phi \rightarrow \phi ^{*}$ and $\chi \rightarrow -\chi ^{*}$. The
parameters were chosen for furnishing just three bounded solutions. Note the
position probability density has a lonely hump for the omnipresent
zero-eigenmode solution. The existence of excited states, though, depends on
the relation between the parameter $\lambda $ and $m$ as given by the top
line of Eq. (\ref{27}).

We have found the solutions of the Dirac equation in 1+1 dimensions for
massive fermions coupled to a time-independent potential. Although the
coupling is of a Lorentz scalar nature we note that due to its coordinate
dependence the potential can not be classified as a true Lorentz scalar.
Nevertheless the analysis carried out here can be illuminating to understand
the conditions under which a scalar potential can hold bound states, the
existence of zero-energy eigenstates as well as its connection to the
nonrelativistic and ultrarelativistic limits of the Dirac equation. We also
note that the potential we have been dealing with always has the zero-mode
eigenstate as a localized solution for $A<0$, regardless the values of the
parameters $A$ and $\lambda $, and it can be considered as the ground-state
solution. On the other hand, the existence of a finite sequence of excited
states depends on the relation between the parameter $\lambda $ and the mass
of the fermion.

Furthermore, the conditions for the existence of a zero-eigenmode for a
general scalar potential, equations (\ref{29d})-(\ref{29g}), including those
ones for the massless case, are more general than that one found by Jackiw
and Rebbi \cite{jackiw}, because we did not restrict ourselves to a
solitonic topological scalar field. These last configurations have been
shown to be responsible for the induction of fractional fermion number on
the vacuum in the second quantization scenario in 1+1 dimensions and the
zero-mode of the corresponding Dirac operator plays a fundamental role on
this phenomenon \cite{niemi}. The fermion number fractionization in quantum
field theory, as well as the role played by the zero-mode and the continuous
spectrum is under investigation by considering the interaction of fermions
with nontopological scalar and pseudoscalar backgrounds and will be reported
elsewhere.\bigskip \bigskip \bigskip

\noindent \textbf{Acknowledgments}

This work was supported in part by means of funds provided by CNPq and
FAPESP.

\newpage

\newpage

\begin{thebibliography}{99}
\bibitem{col1}  S. Coleman, R. Jackiw, L. Susskind, Ann. Phys. (N.Y.) 93
(1975) 267.

\bibitem{col2}  S. Coleman, Ann. Phys. (N.Y.) 101 (1976) 239.

\bibitem{fro}  J. Fr\"{o}hlich, E. Seiler, Helv. Phys. Acta 49 (1976) 889.

\bibitem{cap}  A.Z. Capri, R. Ferrari, Can. J. Phys. 63 (1985) 1029.

\bibitem{gal}  H. Gali\'{c}, Am. J. Phys. 56 (1988) 312.

\bibitem{hoo}  G.\'{}t Hooft, Nucl. Phys. B 75 (1974) 461.

\bibitem{kog}  J. Kogut, L. Susskind, Phys. Rev. D 9 (1974) 3501.

\bibitem{bha}  R.S. Bhalerao, B. Ram, Am. J. Phys. 69 (2001) 817.

\bibitem{cas1}  A.S. de Castro, Am. J. Phys. 70 (2002) 450.

\bibitem{cav}  R.M. Cavalcanti, Am. J. Phys. 70 (2002) 451.

\bibitem{hil}  J.R. Hiller, Am. J. Phys. 70 (2002) 522.

\bibitem{cas2}  A.S. de Castro, Phys. Lett. A 305 (2002) 100.

\bibitem{ntd}  Y. Nogami, F.M. Toyama, W. van Dijk, Am. J. Phys. 71 (2003)
950.

\bibitem{spe}  H.N. Spector, J. Lee, Am. J. Phys. 53 (1985) 248.

\bibitem{mos}  R.E. Moss, Am. J. Phys. 55 (1987) 397.

\bibitem{ho}  C.-L. Ho, V.R. Khalilov, Phys. Rev. D 63 (2001) 027701.

\bibitem{pl}  A.S. de Castro, Phys. Lett. A 328 (2004) 289.

\bibitem{flu}  S. Fl\"{u}gge, Practical Quantum Mechanics, Springer-Verlag,
Berlin, 1999.

\bibitem{haa1}  D. ter Haar, Phys. Rev. 70 (1946) 222.

\bibitem{br}  K.A. Brueckner, Phys. Rev. 103 (1956) 172.

\bibitem{bm}  V. Bernard, C. Mahaux, Phys. Rev. C 23 (1981) 888.

\bibitem{mor1}  P.M. Morse, Phys. Rev. 34 (1929) 57.

\bibitem{mor2}  P.M. Morse, J.B. Fisk, L.I. Schiff, Phys. Rev. 50 (1936) 748.

\bibitem{haa}  D. ter Haar, Problems in Quantum Mechanics, Pion, London,
1975.

\bibitem{nie}  M.M. Nieto, L.M. Simmons Jr., Phys. Rev. A 19 (1979) 438.

\bibitem{ahmed}  Z. Ahmed, Phys. Lett. A 290 (2001) 19.

\bibitem{str}  P. Strange, Relativistic Quantum Mechanics, Cambridge
University Press, Cambridge, 1998.

\bibitem{tha}  B. Thaller, The Dirac Equation, Springer-Verlag, Berlin, 1992.

\bibitem{cn}  F.A.B. Coutinho, Y. Nogami, Phys. Lett. A 124 (1987) 211.

\bibitem{cnt}  F.A.B. Coutinho, Y. Nogami, F.M. Toyama, Am. J. Phys. 56
(1988) 904.

\bibitem{abr}  M. Abramowitz, I.A. Stegun, Handbook of Mathematical
Functions, Dover, Toronto, 1965.

\bibitem{jackiw}  R. Jackiw, C. Rebbi, Phys. Rev. D 13 (1976) 3398.

\bibitem{niemi}  A. Niemi, G. Semenoff, Phys. Rep. 135 (1986) 99.
\end{thebibliography}
\end{document}